\documentclass[twocolumn]{revtex4-1}

\usepackage{latexsym}
\usepackage{amssymb}
\usepackage{amsmath}
\usepackage{amsthm}
\usepackage{amsfonts}
\usepackage{ifthen}
\usepackage{revsymb}
\usepackage{yfonts}
\usepackage{graphicx}
\usepackage{algpseudocode}
\usepackage{bbm}
\usepackage{multirow}
\usepackage{gensymb}
\usepackage{epstopdf}
\usepackage{color}
\usepackage{afterpage}
\usepackage{verbatim}
\usepackage{soul}
\usepackage{lineno, blindtext}
\usepackage{footnote}
\usepackage{afterpage}
\usepackage{algpseudocode}
\usepackage[normalem]{ulem}
\usepackage[bottom]{footmisc}
\usepackage{algorithm}
\usepackage{algpseudocode}
\usepackage[colorlinks = true]{hyperref}
\usepackage{longtable}
\usepackage{xcolor}
\definecolor{darkred}  {rgb}{0.5,0,0}
\definecolor{darkblue} {rgb}{0,0,0.5}
\definecolor{darkgreen}{rgb}{0,0.5,0}

\hypersetup{
  urlcolor   = blue,         % color of external links
  linkcolor  = darkblue,     % color of internal links
  citecolor  = darkgreen,    % color of links to bibliography
  filecolor  = darkred       % color of file links
}

\newcommand{\be}{\begin{equation}}
\newcommand{\ee}{\end{equation}}
\newcommand{\bq}{\begin{eqnarray}}
\newcommand{\eq}{\end{eqnarray}}
\newcommand{\bea}{\begin{eqnarray}}
\newcommand{\eea}{\end{eqnarray}}
\newcommand{\ba}{\begin{align}}
\newcommand{\ea}{\end{align}}

\newcommand{\ket}[1]{ | \, #1 \rangle}
\newcommand{\bra}[1]{ \langle #1 \,  |}

\definecolor{mygray}{gray}{0.6}

\definecolor{mygray}{gray}{0.9}

\begin{document}

\title{Exact search algorithm to factorize large biprimes and a triprime on IBM quantum computer}

\author{Avinash Dash}
\author{Deepankar Sarmah}
\author{Bikash K. Behera}
\author{Prasanta K. Panigrahi }\affiliation{Department of Physical Sciences, Indian Institute of Science Education and Research Kolkata, Mohanpur, 741246, West Bengal, India}

%\date{\today}

\maketitle
\noindent

\textbf{
   Factoring large integers \cite{qfa_BennettNature2000,qfa_KaneNature1998} using a quantum computer is an outstanding research problem that can illustrate true quantum advantage \cite{qfa_BravyarXiv1995,qfa_RisteQIP2017} over classical computers. Exponential time order \cite{qfa_VandersypenNature2001} is required in order to find the prime factors of an integer by means of classical computation. However, the order can be drastically reduced by converting the factorization problem to an optimization one and solving it using a quantum computer \cite{qfa_BurgesMicrosoft2002,qfa_XuPRL2012}. Recent works involving both theoretical and experimental approaches use Shor's algorithm \cite{qfa_VandersypenNature2001,qfa_LopezNatPhot2012,qfa_BocharovPRA2017}, adiabatic quantum computation \cite{qfa_PengPRL2008,qfa_XuPRL2012,qfa_DattaniarXiv2014,qfa_LiarXiv2017,qfa_XuPRL2017} and quantum annealing principles \cite{qfa_DridiScirep2017} to factorize integers. However, our work makes use of the generalized Grover's algorithm as proposed by Liu \cite{qfa_LiuIJTP2014}, with an optimal version of classical algorithm/analytic algebra. We utilize the phase-matching property \cite{qfa_li2007phase} of the above algorithm for only amplitude amplification purposes to avoid an inherent phase factor that prevents perfect implementation of the algorithm. Here we experimentally demonstrate the factorization of two bi-primes, $4088459$ and $966887$ using IBM's $5$- and $16$-qubit quantum processors, hence making those the largest numbers that has been factorized on a quantum device. Using the above $5$-qubit processor, we also realize the factorization of a tri-prime integer $175$, which had not been achieved to date \cite{qfa_DattaniarXiv2014}. We observe good agreement between experimental and theoretical results with high fidelities. The difficulty of the factorization experiments has been analyzed and it has been concluded that the solution to this problem depends on the level of simplification chosen, not the size of the number factored \cite{qfa_SmolinNature2013}. In principle, our results can be extended to factorize any multi-prime integer with minimum quantum resources.
   } 
    
%\section{Introduction}
Prime factorization is one of the NP-class problems that lies at the heart of secure data transmission \cite{qfa_Traversaa2017}. Cryptography techniques such as RSA \cite{qfa_RSAACM1978} have relied on this property to ensure secure means of data communication. It is well-known that the popularly used RSA cryptosystem will be rendered inoperable if the integer factorization problem could be solved in polynomial-time \cite{qfa_WangChina2017}. In $1994$, Shor demonstrated a quantum algorithm \cite{qfa_ShorSIAM1997} which can factorize an integer $N$ in polynomial time-- specifically, it takes quantum gates of order $O((log\,N)^2(log\,log\,N)(log\,log\,log\,N))$ using fast multiplication method. Then in $2001$, Shor's algorithm was experimentally realized using NMR architecture, to factorize $N=15$ \cite{qfa_VandersypenNature2001}. This long standing record was broken in $2012$ with the factorization of $N=21$, which also used Shor's algorithm \cite{qfa_LopezNatPhot2012}. However, these implementations required prior knowledge of the answer \cite{qfa_SmolinNature2013}.

An alternative approach to Shor's algorithm for quantum factorization takes advantage of adiabatic quantum computation \cite{qfa_PengPRL2008}. This approach involves a pre-processing part requiring the transformation of the given factorization problem into an optimization problem \cite{qfa_BurgesMicrosoft2002}, which is reducible to a set of equations by minimization. The set of equations thus formed are used to derive a complex Hamiltonian, which encodes the solution in its ground states. The first number that had been factorized using this technique was $143$, which required only $4$ qubits \cite{qfa_XuPRL2012}. Dattani and Bryans \cite{qfa_DattaniarXiv2014} extended the results to demonstrate factorization of larger numbers and even the tri-prime, $N=175$. However, $175$ had not been experimentally factorized till date, plausibly because the Hamiltonian had been difficult to implement experimentally. A new insight into quantum factorization was presented by Dridi and Alghassi \cite{qfa_DridiScirep2017} recently in the year 2017, where they proposed applying quantum annealing techniques to the same optimization method discussed earlier. The authors showed the experimental factorization of certain bi-primes upto nearly $200000$. In the year $2017$, the integer $N=291311$ had been factorized using the adiabatic approach, making it the largest number that has been factorized using a quantum device \cite{qfa_LiarXiv2017}. Only $3$ qubits were used in this case, as it was observed that further minimization could be achieved during the pre-processing part. 

In the course of our current work, the method of minimization has been followed for requisite pre-processing, as is required in the adiabatic approach. However, instead of dealing with a dynamically evolving system Hamiltonian, we directly implement an unitary operation which is an exponential function of the non-unitary Hamiltonian used in the adiabatic case. The basis states encoding the required solutions could then be separated out using an exact quantum search algorithm \cite{qfa_LiuIJTP2014}. Implementing this protocol on IBM's quantum processors (both $5$-qubit and $16$-qubit), we have experimentally factorized the integers $4088459$ and $966887$, using $2$ and $4$ qubits respectively. The choice of these numbers has been made in order to show that the complexity of the factorization problem (via this approach) does not depend on the largeness of the number, but is rather dependent on a certain property of the factors. Apart from these bi-prime numbers, we have also demonstrated the experimental factorization of a tri-prime $N=175$ which requires only $2$ qubits. 

%\section{Results}
%\label{qfa_results}
%\subsection{Quantum Factorization of large numbers having 2 factors}

In general, we seek to find out the prime factors $p$ and $q$ of an odd composite number $N$, i.e. $N=p \times q$. If the number to be factorized is even, then we keep dividing it by two until an odd composite number is reached. The binary form of the factors $p$ and $q$ can be denoted as $\{1p_mp_{m-1}...p_2p_11\}_{bin}$ and $\{1q_nq_{n-1}...q_2q_11\}_{bin}$ respectively, where $p= 2^{m+1}+\sum_{i=1}^{i=m}p_i2^i+1$ and the same expression applies for $q$. Our aim is to first optimize the factorization problem into a set of equations in variables $\{p_i\}$ and $\{q_j\}$ ($i\in [1,m]\cap \mathbb{N}, j\in [1,n]\cap \mathbb{N}$), subject to the condition that $p_{bin}\times q_{bin}= N_{bin}$ (the subscript denotes binary notation) \cite{qfa_BurgesMicrosoft2002,qfa_XuPRL2012,qfa_DattaniarXiv2014,qfa_LiarXiv2017}. It has been recently shown that the set of equations in $m+n$ variables is reducible to a smaller number of variables \cite{qfa_DattaniarXiv2014}. This is because, if $p_i+q_j= z$ for some $i,j$, where $z\in \{0,2\}$, then $p_i= q_j= 0$ if $z= 0$ and $p_i= q_j= 1$ if $z=2$. Hence, the set of equations is reduced to only those variables $\{p_i\}$ and $\{q_j\}$ such that the index $i$ takes values from the set $I_p=\{i\in [1,m]\cap\mathbb{N}: p_i+q_j=1\:for\:some\:j\}$ and same works for index $j$ also. In cases where $m= n$, the optimization of the set of equations in variables $\{p_i\}$ and $\{q_i\}$ leads to a smaller set of equations where $p_i+ q_i= 1$ and $\sum_{i< j} p_iq_j+p_jq_i= z$ ($z\in \{0,1\}, i,j\in I\subseteq [1,n]\cap\mathbb{N}$). Hence, the set $I$ such that $p_i+q_i=1$ and $p_i\neq q_i$ $\:\forall i\in I$.

Consider $N= 4088459$. It is known that the prime factors of this number have the same number of digits. Hence, $m= n$ in this case. The prime factors $p$ and $q$ are denoted in binary as $\{1p_9p_8...p_2p_11\}_{bin}$ and $\{1q_9q_8...q_2q_11\}_{bin}$ respectively. The factorization problem reduces to the set of equations:

\begin{eqnarray}
        p_1+q_1=1 && \nonumber\\
        p_3+q_3=1 && \nonumber\\
        p_1q_3+p_3q_1=0 &&
       \label{qfa_eq_1}
\end{eqnarray}

As for the rest of the variables, upon optimization we obtain:
\begin{eqnarray}
       && p_i=q_i=0\text{; }i\in \{2,4\} \nonumber\\
       && p_i=q_i=1\text{; }i\in \{5,6,7,8,9\}
       \label{qfa_eq_2}
\end{eqnarray}

Since $q_i=1-p_i$ for $i\in \{1,3\}$, the set of equations (Eq.~\eqref{qfa_eq_1}) is further reduced to
\begin{equation}
    q_1+q_3-2q_1q_3=0
    \label{qfa_eq_3}
\end{equation}

The values of $q_1$ and $q_3$ satisfying Eq.~\eqref{qfa_eq_3}, represents the solution to our factorization problem that are encoded in the ground state of the $2$-qubit Hamiltonian,

\begin{equation}
    \hat{H}= (\hat{a_1}+\hat{a_2}-2\hat{a_1}\hat{a_2})^2
    \label{qfa_eq_4}
\end{equation}

where $\hat{a_i}=\frac{I-\sigma_z^i}{2}$, $I$ stands for the $1$-qubit identity operation and $\sigma_z^i$ denotes the Pauli $Z$ operator acting on the $i$th qubit. Since $q_1$, $q_3$ satisfy Eq.~\eqref{qfa_eq_3}, the two qubit $z$-basis eigenstate $\ket{q_1}\ket{q_3}$ satisfies $H\ket{q_1}\ket{q_3}=0.\ket{q_1}\ket{q_3}$ (note that for any $b\in \{0,1\}$, $\hat{a_i}\ket{b}=b\ket{b}$, thus yielding the above result), while for any other two qubit state (in $z$-basis) the corresponding eigenvalue of $\hat{H}$ is some positive value (non-zero). It is to be noted that, for a case in which two factors need to be found, $\hat{H}$ can have two and only two ground state eigenstates (whose eigenvalue is zero). Upon simplifying, we obtain

\begin{equation}
    \hat{H}= \frac{1}{2}(I_2-\sigma_z^1\otimes\sigma_z^2)
    \label{qfa_eq_5}
\end{equation}

where we have used the fact that $\hat{a_i}^2=\hat{a_i}$. $I_2$ denotes the $2$-qubit identity operation. It can be shown that $\hat{H}$ has eigenvalues $0$ and $1$.

\begin{figure*}
\centering
\includegraphics[scale=0.8]{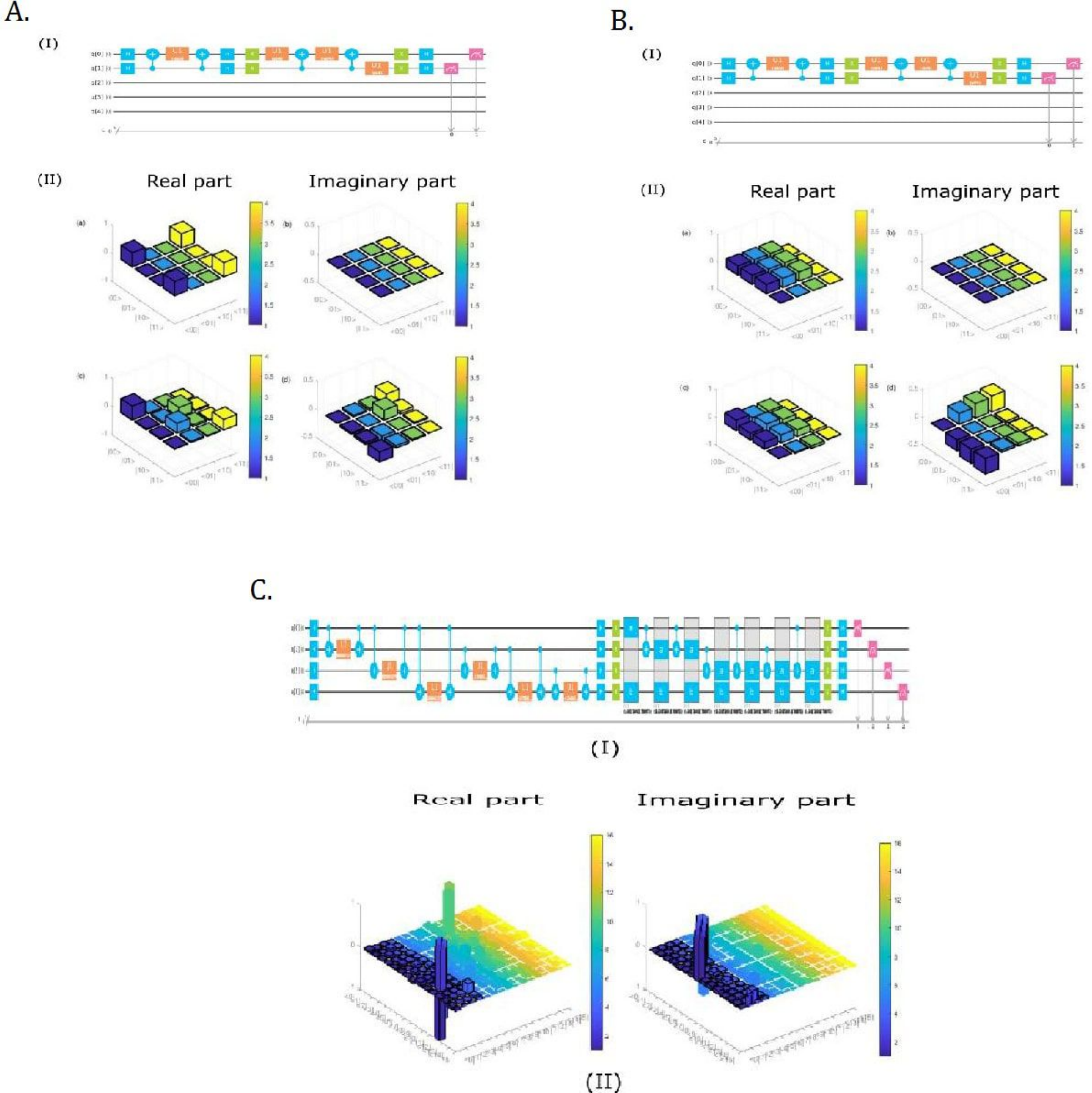}
\caption{\textbf{Experimental implementation with quantum circuits and results.} The circuits \textbf{A (I)} and \textbf{B (I)} were implemented using IBM's $5$-qubit quantum processor `ibmqx4' to explicitly solve the factorization problem for $N=4088459$ and $N=175$ respectively using only $2$ qubits. A classical simulation was used in case of \textbf{C} to factorize $N=966887$. The experiments and the simulation were performed $8192$ times. The channels used are such that gate errors are minimized.  Measurements are performed in $Z$-basis. To check the accuracy of our experimental results, quantum state tomography is performed in each case as illustrated by Figs. \textbf{A (II)}, \textbf{B (II)} and \textbf{C (III)}. Real (left) and imaginary (right) parts of the reconstructed theoretical (\textbf{A (II (a,b)), B(II (a,b))}) and experimental (\textbf{A (II (c,d)), B (II (c,d))}) and simulational (\textbf{C (II)}) density matrices have been presented for the experiments conducted for $N=4088459$, $N=175$ and $N=966887$ respectively. \textbf{A} Our experimental results indicate that the states $\ket{00}$ and $\ket{11}$ appear with a high probability acting as the ground states of the two qubit Hamiltonian $\hat{H}$ (Eq. \eqref{qfa_eq_5}). As a result, $(q_1,q_3)=(0,0)$ and $(q_1,q_3)=(1,1)$ are the solutions (Eq. \eqref{qfa_eq_3}) to the factorization of $N=4088459$ that corresponds to $2017$ and $2027$ in decimal respectively. The fidelity of the results is found to be 0.9278. \textbf{B. } The experimental result shows that the states $\ket{00}$, $\ket{01}$ and $\ket{10}$ are obtained with nearly equal probabilities, hence acts as the solution states. Here, {($p_1$,$q_1$)} $\in$ $\{(0,0),(0,1),(1,0)\}$ are the solutions (Eq. \eqref{qfa_eq_13}) to the problem that corresponds to the factors of $175$ as $5$, $5$ and $7$ in decimal respectively. The fidelity of this case is found to be 0.9850. \textbf{C.} The simulation result shows that the states $\ket{0111}$ and $\ket{1000}$ are obtained with nearly equal probabilities representing as the solution states that correspond to the factors $(q_1,q_2,q_3,q_6)\in \{(0,1,1,1),(1,0,0,0)\}$ (Eq. \ref{qfa_eq_10}), in decimal $1021$ and $947$ respectively.}
\label{qfa_Fig1}
\end{figure*}

One can verify that the unitary operator $e^{-i\hat{H}\theta}$ (equivalent to the operation ($e^{-i\theta}-1$)$\hat{H}+I$) induces a relative phase change of $\theta$ in the $2$-qubit $z$-basis eigenstates that correspond to the eigenvalue $0$ for the operator $\hat{H}$. If $\ket{b_1}\ket{b_2}$ is such a state ($b_1,b_2\in \{0,1\}$), then $(q_1,q_3)=(b_1,b_2)$. Hence, the operator $e^{-i\hat{H}\theta}$ performs a conditional phase shift $e^{i\theta}$ which marks the required ``solution" states.
Firstly, we take the equal superposition state in a two qubit system, i.e. the state given by $\ket{\psi_0}=\frac{1}{2}\sum_{i=0}^{i=3}\ket{i}$. We pass $\ket{\psi_0}$ through the operator $e^{-i\hat{H}\theta}$, which marks our required solution states. Our next aim is to separate out these ``marked" states which we wish to obtain by some means. To achieve this, the generalized Grover's search algorithm is used, which searches any number of marked states from an arbitrary quantum database with certainty \cite{qfa_LiuIJTP2014}. We have already achieved the first step of this algorithm, viz. we have introduced a conditional phase shift to the marked states. Secondly, we apply the 2 qubit operator $U^\dag$, where $U$ transforms $\ket{00}$ to $\ket{\psi_0}$. In our case, $U^\dag=U=H^{\otimes2}$ ($H$ denotes the Hadamard operation). The next step involves the conditional phase shift $e^{i\theta}$ to $\ket{00}$ state, whereas all the other basis states remain unchanged. Finally, we perform the operation $U$ (in our case $U=H^{\otimes2}$. The result is that the final state should contain only the solution states. The value of the phase shift angle $\theta$ for the exact quantum search algorithm to work is found to be equal to $\frac{\pi}{2}$ \cite{qfa_LiuIJTP2014}, with only a single iteration being required. Detailed calculations along with the overall quantum circuit have been presented in Supplementary material. We experimentally realized the factorization problem at hand using IBM's $5$-qubit quantum processor. The tomographical results have been presented in Fig.~\ref{qfa_Fig1} \textbf{A}.

It is pointed out that a two qubit Hamiltonian was used to factorize $N=4088459$. This was because the equation obtained after simplification (Eqs.~\eqref{qfa_eq_3}) involved two variables. These variables were of the form $q_i$ such that $p_i\neq q_i$. The values of all the other variables were resolved during the simplification process. Hence, from the above observation, we may conclude that, if the two prime factors (having same number of digits in binary notation) of an odd composite number differ at $n$ bits, the factorization problem is solvable by using $n$ qubits. The solution is encoded in the $n$ ground states of the $n$ qubit Hamiltonian. Let us quickly take another example in which a 2 qubit Hamiltonian is required to factorize the number. Consider $N=143$. In this case, the two factors are $11$ and $13$, or correspondingly $\{1011\}_{bin}$ and $\{1101\}_{bin}$ respectively. Since the factors differ at 2 bits, two qubits are required to factorize $143$, with the expected ``solution" states (ground states of the formed Hamiltonian) to be $\ket{01}$ and $\ket{10}$. The quantum circuit remains the same in this case, as it does for all cases where 2 qubits are required for factorization via this method. The only difference is that, upon evaluating the Hamiltonian, the operation $e^{-i\hat{H}\theta}$ is decomposed into $e^{-i(\sigma_z^1\otimes\sigma_z^2)\frac{\theta}{2}}$ instead of $e^{i(\sigma_z^1\otimes\sigma_z^2)\frac{\theta}{2}}$, which conditionally adds $\theta$ phase to $\ket{01}$ and $\ket{10}$ states instead of $\ket{00}$ and $\ket{11}$ states as previously. 

Now we implement our protocol for the factorization of $N=966887$. The two prime factors $p$ and $q$ are denoted as $\{1p_8p_7...p_2p_11\}_{bin}$ and $\{1q_8q_7...q_2q_11\}_{bin}$ respectively. The simplification of the crude set of equations results in the following set of equations;

\begin{eqnarray}
       && p_i+q_i=1\text{; }i\in \{1,2,3,6\} \nonumber \\
       && p_iq_1+q_ip_1-1=0\text{; }i\in\{2,3,6\}\nonumber \\
       && p_iq_j+p_jq_i=0\text{; }2\leq i<j\leq 6\text{; }i,j\in \{2,3,6\}
       \label{qfa_eq_10}
\end{eqnarray}

The rest of the variables turn out to be
\begin{equation}
    p_i=q_i=1\text{; }i\in \{4,5,7,8\}
    \label{qfa_eq_23}
\end{equation}

We map the variables ($q_1,q_2,q_3,q_6$) to the operators ($\hat{a_1},\hat{a_2},\hat{a_3},\hat{a_4}$), where $\{\hat{a_i}\}$ are chosen in the same manner as previously. As a result, the four qubit Hamiltonian obtained in this case is 
\begin{eqnarray}
    \hat{H} = \frac{1}{2}[ && 6I_4+\sigma_z^1\otimes\sigma_z^2\otimes I\otimes I+\sigma_z^1\otimes I\otimes\sigma_z^3\otimes I\nonumber \\
    && +\sigma_z^1\otimes I\otimes I\otimes\sigma_z^4-I\otimes\sigma_z^2\otimes\sigma_z^3\otimes I \nonumber \\ &&-I\otimes\sigma_z^2\otimes I\otimes\sigma_z^4-I\otimes I\otimes\sigma_z^3\otimes\sigma_z^4]
    \label{qfa_eq_11}
\end{eqnarray}

where $I_4$ is the $4$-qubit identity operation.

In such a four qubit case, implementing the $e^{-i\hat{H}\theta}$ operation introduces a relative phase shift of $3\theta$ to the solution states w.r.t. \textit{some} of the basis states (one of them being $\ket{0000}$). Again, since there are two factors, there will be two ground state eigenstates of $\hat{H}$. We separate these solution states from the other states using the exact quantum search algorithm as above. In this case, the quantum database state $\ket{\psi_0}$ is the $4$-qubit equal superposition state.

Like previously, this case also requires only $1$ iteration of the quantum search algorithm. The rest of the protocol is followed in the same manner. It must be noted that, here a conditional phase shift $e^{i3\theta}$ must be applied to $\ket{0000}$ state. 

The circuit implementation has been described in Supplementary Section. We have used IBM's classical topology to simulate the problem. The simulational results are shown in Fig.~\ref{qfa_Fig1} \textbf{C}.

Theoretical evaluation yields the following phase shifts (Table~\ref{qfa_table1}) induced by the $e^{-i\hat{H}\theta}$ operation on the four qubit $z$-basis states. All phase shifts are relative to $\ket{0000}$ state.

\begin{table}[H]
\centering
\caption{\emph{$z$-basis eigenstates and the phase shift (relative to $\ket{0000}$) induced on them by the operation $e^{-i\hat{H}\theta}$.}}
\begin{tabular}{cccc}
\hline
\hline
z-basis state  & Phase shift & z-basis state & Phase shift \\ \hline
$\ket{0000}$   &  $0$        & $\ket{1000}$  & $3\theta$   \\
$\ket{0001}$   & $-\theta$   & $\ket{1001}$  & $0$         \\ 
$\ket{0010}$   & $-\theta$   & $\ket{1010}$  & $0$         \\
$\ket{0011}$   & $0$         & $\ket{1011}$  & $-\theta$   \\
$\ket{0100}$   & $-\theta$   & $\ket{1100}$  & $0$         \\ 
$\ket{0101}$   & $0$         & $\ket{1101}$  & $-\theta$   \\
$\ket{0110}$   & $0$         & $\ket{1110}$  & $-\theta$   \\
$\ket{0111}$   & $3\theta$   & $\ket{1111}$  & $0$         \\
\hline
\hline
\end{tabular}
\label{qfa_table1}
\end{table}

It is observed that, although a phase of $3\theta$ is introduced to two of the basis states, which are supposedly our solution states, there exist $6$ other basis states which also have some phase ($-\theta$) added to them. On putting ($q_1,q_2,q_3,q_6$)$=$($1,0,0,0$) or ($0,1,1,1$), we obtain $q=947$ or $1021$. It is indeed true that $966887=1021\times947$. Hence, apart from a conditional phase shift to the two solution states (ground states of $\hat{H}$), the $e^{-i\hat{H}\theta}$ operation for a $4$-qubit Hamiltonian also incorporates a conditional phase shift to certain other basis states through a negative angle of equal magnitude. This phenomenon is not observed in case of a two or three qubit Hamiltonian. In the previous example of $N=4088459$, in which a $2$-qubit Hamiltonian was used for factorization, the basis states other than the solution states exhibited no phase shift relative to the $2$-qubit $\ket{0}$ state upon the application of $e^{-i\hat{H}\theta}$. The same is observed for cases in which the two prime factors happen to differ at $3$ bits, viz. a $3$-qubit Hamiltonian is required. The $4$-qubit Hamiltonian has different eigenvalues for basis states other than the two ground states. For some basis states, it has the eigenvalue $3$, resulting in a relative phase shift of $0$ while for others it has the eigenvalue $4$, corresponding to a relative phase shift of $-\theta$ (Table~\ref{qfa_table1}). This trend is expected to continue for $n>4$. Despite the appearance of unwanted phases (of less magnitude) on non-solution states, the exact quantum search algorithm can still be used. Because of the phase-matching property of the algorithm, the ``marked" solution states are nevertheless amplified as a result. Hence, the solution states are obtained with a significantly higher probability than the other basis states. This result of the generalized Grover's algorithm can be thought of as an analogy to resonance phenomena.

The circuit for factorizing $N=966887$ can be used to factorize all numbers whose two prime factors happen to differ at $4$ bits. The only difference will be in the signs of the phase angles used in the unitary implementation of the $e^{-i\hat{H}\theta}$ operation. 

%\subsection{Quantum factorization of a number with 3 prime factors}

We take $N=175$, with three prime factors to show how the protocol works in this case. Here $N =p\times q \times r$, where $p$, $q$ and $r$ are denoted as $\{1p_11\}_{bin}$, $\{1q_11\}_{bin}$ and $\{1r_11\}_{bin}$ respectively. The following set of equations are obtained upon simplification:
 \begin{eqnarray}
    && p_1 + q_1 + r_1 = 1\nonumber \\
    && p_1q_1 + q_1r_1 + p_1r_1 = 0 \nonumber \\
    \implies && p_1+q_1-p_1^2-q_1^2 - p_1q_1 = 0
    \label{qfa_eq_13}
\end{eqnarray}
 Mapping variables $(p_1,q_1)$ to the operators $(\hat{a_1},\hat{a_2})$, the corresponding Hamiltonian is evaluated as
 \begin{equation}
        \hat{H} = \hat{a_1}\hat{a_2} = \frac{1}{4}\left[I_{2}-\sigma_z^1\otimes I-I\otimes\sigma_z^2+\sigma_z^1\otimes\sigma_z^2\right]
          \label{qfa_eq_14}
 \end{equation}
 
which has the solution states encoded as its ground states.

Hence, $2$ qubits are needed to factorize this number. The Hamiltonian has three ground states, instead of two, since we expect three factors here, of which not all are equal (since $p_1\neq q_1\neq r_1$ is not possible). As a result, the generalized search algorithm must work in such a way such that it searches $3$ marked states from the $2$-qubit equal superposition state. The $e^{-i\hat{H}\theta}$ operation induces a conditional phase shift of $\theta$ in this case. The phase shift angle $\theta=2sin^{-1}\left(\frac{sin\frac{\pi}{6}}{\frac{\sqrt{3}}{2}}\right)$ meets the criteria for the quantum search algorithm to be applicable in this case. Minimum number of iterations required is $1$. The experiment for the factorization of $175$ was performed using IBM's $5$-qubit quantum processor and the results are presented in Fig.~\ref{qfa_Fig1} \textbf{B}. 

We utilized the method of minimization proposed by Burges \cite{qfa_BurgesMicrosoft2002} to factorize the largest numbers on a quantum device as of yet. The method of minimization was useful in an approach of adiabatic quantum computation which had been in use in recent times to factorize bi-prime numbers in experimental systems. Some of the numbers factorized using adiabatic principles are $143$ \cite{qfa_XuPRL2012}, $56153$ \cite{qfa_DattaniarXiv2014} and $291311$ \cite{qfa_LiarXiv2017}. A recent work by Dridi and Alghassi \cite{qfa_DridiScirep2017} presented an autonomous algorithm to factorize bi-prime numbers using the technique of quantum annealing. In our paper, we have neither involved a dynamically evolving Hamiltonian characteristic of the adiabatic approach, nor have we attempted to search for the global minimum from a large set of Hamiltonians, which is a feature of the annealing approach. Instead, we have put to use an exponential function of the Hamiltonian derived from the minimization method, which is found to conditionally mark the ``solution" states with a certain (equal) phase. To separate out these states, we utilized the generalized Grover's algorithm proposed by Liu \cite{qfa_LiuIJTP2014}, which is an exact quantum search algorithm. Using this technique, we were able to factorize the numbers $4088459$, $966887$ and $175$, the first two being the largest bi-primes to date and the last being the first tri-prime that have been factorized on a quantum device. We showed that the number of qubits required to factorize any given bi-prime number is equal to the number of bits at which the binary notations of the two factors differ. As a result, we were able to factorize $4088459$ using only $2$ qubits. Interestingly, it was noted that to factorize a smaller number like $966887$, more number of qubits as well as a lengthier computation were required. However, some inherent errors were also observed in this case, as a result of the $4$-qubit Hamitonian exhibiting two different eigenvalues other than zero. However, this issue was resolved by using the phase-matching property of the search algorithm, which resulted in the amplification of our desired states. Finally, a major challenge before us is the pre-processing part leading to the formation of the Hamiltonian, which has to be carried out by a computer program.

{\bf Supplementary Information} is available in the online version of this paper.

{\bf Acknowledgments}
We thank Manabputra and D.T. Rao for experimental contributions; P. Krishna for assistance with the experimental control software. We acknowledge the support of IBM Quantum Experience for providing access to the quantum processors. The views expressed are those of the authors and do not reflect the official policy or position of IBM or the IBM Quantum Experience team. We acknowledge financial support from Kishore Vaigyanik Protsahan Yojana (KVPY) and Department of Science and Technology (DST), Government of India.

{\bf Author contributions}
A.D. developed the protocol and designed the quantum circuits. D.S. and A.D. drew the circuits on IBM quantum experience platform and performed the experiments. D.S. interpreted and analyzed the experimental data. A.D., B.K.B. and D.S. contributed to the composition of the manuscript. P.K.P. thoroughly checked and reviewed the manuscript.

{\bf Author information} The authors declare no competing financial interests. Correspondence and requests for materials should be addressed to P.K.P. (pprasanta@iiserkol.ac.in).

\onecolumngrid
\section*{Supplementary Information: Exact search algorithm to factorize large biprimes and a triprime on IBM quantum computer}
%\label{qfa_methods}
\textbf{Circuit used for the factorization of $N=4088459$}
\begin{figure}[H]
\includegraphics[width=\linewidth]{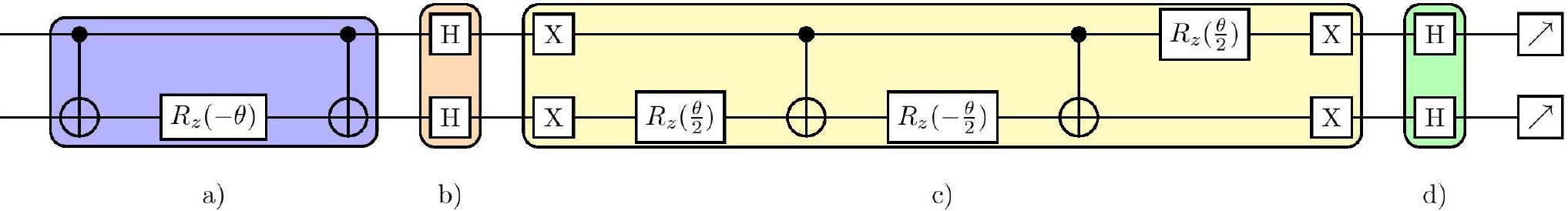}
\caption{\textbf{The circuit implementation for factorizing $N=4088459$.} The operation $e^{-i\hat{H}\theta}$ can be expressed as $e^{-i\hat{H}\theta}=e^{-iI_{2}\frac{\theta}{2}}e^{i(\sigma_z^1\otimes\sigma_z^2)\frac{\theta}{2}}$. Since the operation $e^{-iI_{2}\frac{\theta}{2}}$ serves no other purpose than to introduce a global phase of $\frac{\theta}{2}$ to the system, it carries no physical significance. Hence, to simulate the effect of the $e^{-i\hat{H}\theta}$ on the system (introduction of a conditional phase shift), we simply have to implement the $e^{i(\sigma_z^1\otimes\sigma_z^2)\frac{\theta}{2}}$ operation. \textbf{(a)} The operation $e^{i(\sigma_z^1\otimes\sigma_z^2)\frac{\theta}{2}}$, which is an exponential function of $\sigma_z^1\otimes\sigma_z^2$, can be expressed as the product of certain unitary gates \cite{qfa_sup_WhitfieldarXiv2010}. This is the corresponding circuit implementation. $R_z(-\theta)$ is the rotation operation through an angle $-\theta$ about the $z$-axis of the Bloch sphere. This operation introduces a phase shift $e^{-i\theta}$ to the qubit if and only if it is in $\ket{1}$ state. Overall, the action of this part of the circuit is that it conditionally induces a phase shift of $\theta$ angle only to the ground states of $\hat{H}$, which encode the solution to our problem, relative to the other basis states. Hence, our required ``solution" states have been marked as required for the quantum search algorithm. \textbf{(b)} Since our initial quantum database $\ket{\psi_0}$ from which the marked states are to be searched is taken to be the $2$-qubit equal superposition state, the operation $U$ such that $U\ket{00}=\ket{\psi_0}$ is given by $U=H^{\otimes2}$. In this step, we apply the $U^{\dag}$ operation, which is the corresponding inverse operation. In our case, $U^{\dag}$ is equal to $H^{\otimes2}$ as well. \textbf{(c)} This part of the circuit is for the purpose of implementing the conditional phase shift $e^{i\theta}$ to $\ket{00}$ state only, while the other basis states are left unchanged. \textbf{(d)} Finally, we apply $U=H^{\otimes2}$ to obtain an equal superposition of the marked states only. The value of $\theta$ must be equal to $\frac{\pi}{2}$ for the exact search algorithm to work. The necessary calculation for obtaining this has been presented below. }
\label{qfa_sup_Fig1}
\end{figure}

Fig.~\ref{qfa_sup_Fig1} presents the overall circuit for the given factorization problem. Let us present the necessary formalism for obtaining the value of the phase shift angle $\theta$ for the exact quantum search algorithm to work. Since there are two different factors, the Hamiltonian $\hat{H}$ (Eq.~\eqref{qfa_eq_5}) has two and only two eigenstates whose corresponding eigenvalue is zero. Hence, two basis states represent the solution to our problem. These two states are marked with the relative phase shift $e^{i\theta}$ in the scheme mentioned above. Suppose from the quantum database $\ket{\psi_0}$, $\ket{x_0}$ represents the normalized sum over the two marked states. Let $\ket{x^\perp_{0}}$ represent the normalized sum over the other two states (non-marked states). Hence, in the bidimensional vector space, $\ket{x^\perp_{0}}$ represents the hyperplane perpendicular to $\ket{x_0}$, and the vector space is spanned by $\{\ket{x_0}, \ket{x^\perp_{0}}\}$. The state $\ket{\psi_0}$ can be expressed as

\begin{equation}
    \ket{\psi_0}=sin\phi\ket{x_0}+cos\phi\ket{x^\perp_{0}}
    \label{qfa_eq_6}
\end{equation}

In our case, $\phi=\frac{\pi}{4}$. The relation between $\theta$ and $\phi$ is given as \cite{qfa_sup_LiuIJTP2014}
\begin{equation}
    \theta=2sin^{-1}\left(\frac{sin\frac{\pi}{4j+2}}{sin\phi}\right)
    \label{qfa_eq_7}
\end{equation}

where $j$ is the minimum number of iterations after which the marked states can be separated with certainty. Eq.~\eqref{qfa_eq_7} has real solutions for 
\begin{equation}
    j\geq\frac{\pi}{4\phi}-\frac{1}{2}
    \label{qfa_eq_8}
\end{equation}

Hence, the value of $j$ is given as \cite{qfa_sup_LiuIJTP2014}
\begin{equation}
       j=
       \begin{cases}
       \frac{\pi}{4\phi}-\frac{1}{2}\text{, if }\left(\frac{\pi}{4\phi}-\frac{1}{2}\right) \text{ is an integer}\\ 
       INT[\frac{\pi}{4\phi}-\frac{1}{2}]+1\text{, otherwise}
       \end{cases}
       \label{qfa_eq_9}
\end{equation}

%\subsection{Implementation in IBM Quantum Computer}

The formulation of the theoretical and experimental density matrices for the purpose of carrying out quantum state tomography shall now be discussed. For a two qubit system, the experimental density matrix is given by
\begin{equation}
    \rho^{E}=\frac{1}{4}\sum_{i,j=0}^{3}(S_i\times S_j)(\sigma_i\otimes\sigma_j)
    \label{qfa_eq_16}
\end{equation}

$S_0$, $S_1$, $S_2$ and $S_3$ are known as Stokes parameters. $\{\sigma_i\}$ is the set of single qubit operations $I,\sigma_X,\sigma_Y,\sigma_Z$ for $i=0,1,2,3$ respectively. The Stokes parameters are calculated as
\begin{eqnarray}
    && S_0=1 \nonumber \\
    && S_1=P_{\ket{0_X}}-P_{\ket{1_X}} \nonumber \\
    && S_2=P_{\ket{0_Y}}-P_{\ket{1_Y}} \nonumber \\
    && S_3=P_{\ket{0_Z}}-P_{\ket{1_Z}}
    \label{qfa_eq_17}
\end{eqnarray}

where $P_{\ket{i_j}}$ denotes the probability of obtaining the eigenstate $\ket{i}$ upon measurement in the basis denoted by $j$. To perform a measurement on any qubit in $X$-basis, $H$ gate is applied to the qubit before measurement and $S^{\dag}H$ gate is used for the same in $Y$-basis. Our task is to check whether the experimental density matrix is in good agreement with the theoretical one. The theoretical density matrix is given by
\begin{equation}
    \rho^{T}=\ket{\Psi}\bra{\Psi}
    \label{qfa_eq_18}
\end{equation}

A measure of the overlap between two density matrices is given by fidelity, which quantifies the closeness of the experimentally obtained quantum states to the final state of the system in the ideal case. This quantity is calculated as
\begin{equation}
    F(\rho^T,\rho^E)=Tr\left(\sqrt{\sqrt{\rho^T}\rho^E\sqrt{\rho^T}}\right)
\end{equation}

In theory, the final state obtained after the circuit shown in Fig.~\ref{qfa_sup_Fig1} is executed should be $\ket{\Psi}=\frac{1}{\sqrt{2}}\left[\ket{00}+\ket{11}\right]$. Hence,
\begin{equation}
    \rho^{T}= \frac{1}{2} \left[ {\begin{array}{cccc}
        1 & 0 & 0 & 1  \\
        0 & 0 & 0 & 0 \\
        0 & 0 & 0 & 0 \\
        1 & 0 & 0 & 1 \\
    \end{array} }\right]
    \label{qfa_eq_19}
\end{equation}

After performing the experiment $8192$ times in each basis (given by Eq.~\eqref{qfa_eq_16}), we obtained $\{P_{\ket{i_j}}\}$ and calculated the Stokes parameters. The experimental density matrix was obtained as follows
\begin{equation}
   \rho^{E}= 
  \left[ {\begin{array}{cccc}
        0.4930 & 0.0270 & 0.0330 & -0.0092  \\
        0.0270 & 0.0710 & 0.3393 & 0.0420 \\
        0.0330 & 03393 & 0.0640 & 0.0275 \\
        -0.0092 & 0.0420 & 0.0275 & 0.3720 \\
  \end{array} } \right]+ i\left[ {\begin{array}{cccc}
   0.0000 & -0.0265 & -0.0375 & -0.1820\\
   -0.0265 & 0.0000 & -0.1785 & -0.0435 \\
   -0.0375 & 0.1785 & 0.0000 & -0.0210\\
   0.1820 & 0.0435 & 0.0210 & 0.0000\\
  \end{array} } \right]
  \label{qfa_eq_20}
\end{equation}

\textbf{Circuit used for factorization of $N=966887$}

\begin{figure}[H]
\includegraphics[width=\linewidth]{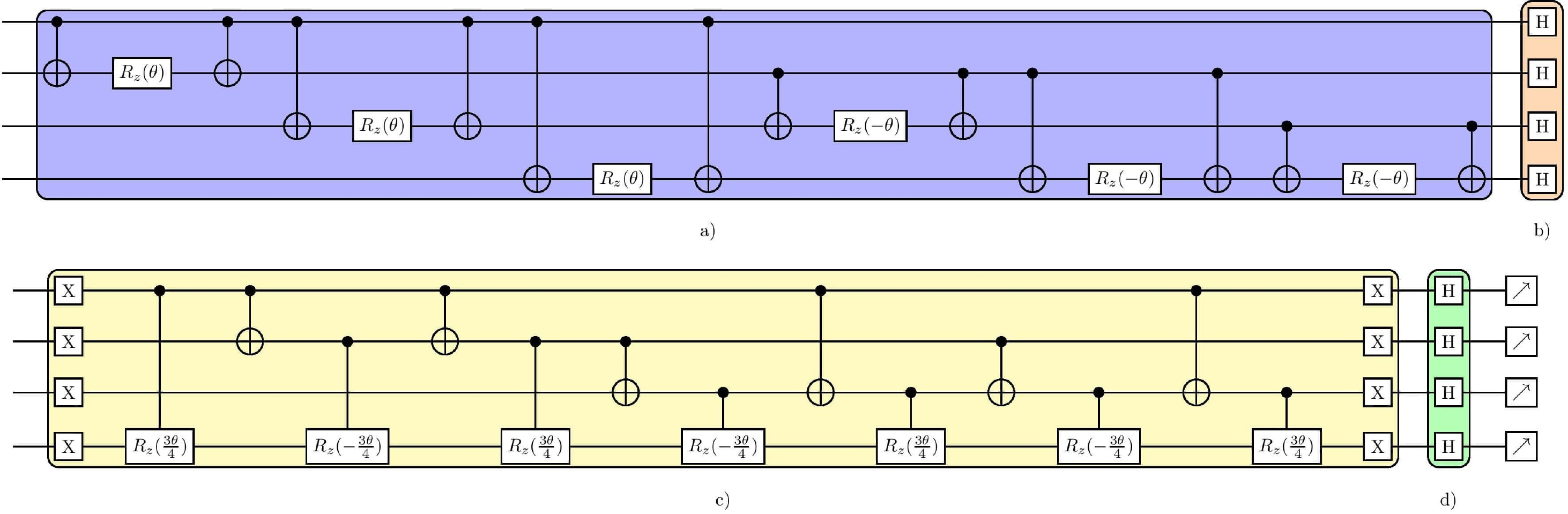}
\caption{\textbf{The circuit implementation of the protocol required for the factorization of $N=966887$.} \textbf{(a)} This is an implementation of $e^{-i\hat{H}\theta}$ operation, where $\hat{H}$ is given by Eq.~\eqref{qfa_eq_11}. The $e^{-3iI_4\theta}$ factor has been neglected. Overall, the action of this part of the circuit is that it conditionally induces an equal phase shift of either $3\theta$ or $4\theta$, depending on the states relative to which the phases are considered, to the ground states of $\hat{H}$. Hence, our required ``solution" states have been marked as required for the quantum search algorithm. \textbf{(b)} Since our initial quantum database $\ket{\psi_0}$ from which the marked states are to be searched is taken to be the $4$-qubit equal superposition state, the operation $U$ such that $U\ket{00}=\ket{\psi_0}$ is given by $U=H^{\otimes4}$. In this step, we apply the $U^{\dag}$ operation, which is the corresponding inverse operation. In our case, $U^{\dag}$ is equal to $H^{\otimes4}$ as well. \textbf{(c)} This part of the circuit is for the purpose of implementing the conditional phase shift $e^{i3\theta}$ to the $4$-qubit $\ket{0}$ state only, while the other basis states are left unchanged. \textbf{(d)} Finally, we apply $U=H^{\otimes4}$ to obtain the final state. It is expected that the solution states shall be obtained with high probabilities upon measurement of the final state. The value of $\theta$ must be chosen according to Eq.~\eqref{qfa_eq_12}. The search algorithm, though not exact in this case, is still suitable for our purpose. Amplitude amplification is still effectuated if appropriate phases are applied whenever required. }
\label{qfa_sup_Fig2}
\end{figure}

Fig.~\ref{qfa_sup_Fig2} presents the overall circuit for the given factorization problem. Following from Eq~\eqref{qfa_eq_7}, the phase shift angle (of the solution states) with respect to some non-solution basis state, say $\ket{0000}$, must satisfy the following criteria. Note that the phase shift angle is $3\theta$ in this case, relative to $\ket{0000}$. 
\begin{equation}
    3\theta=2sin^{-1}\left(\frac{sin\frac{\pi}{6}}{\frac{1}{2\sqrt{2}}}\right)
    \label{qfa_eq_12}
\end{equation}

The factorization problem was simulated using IBM's classical topology. The resultant density matrix was obtained as follows
%\subsection{Implementation in IBM Quantum Computer}
\begin{eqnarray}
 \rho^{S}= \left[{\begin{array}{c|cccccccc}
    &C_{1}&C_{2}&C_{3}&C_{4}&C_{5}&C_{6}&C_{7}&C_{8}\\
    \hline
     R_1&0.0420&0.0080&0.0100& -0.0030&0.0085&0.0028&0.0025&-0.0536\\
        R_2&0.0080&0.0060&0.0490&0.0095&0.0442&0.0085&0.0506&0.0035\\ 
        R_3&0.0100&0.0490&0.0070&0.0090&0.0480&0.0524&0.0075&0.0088\\ 
        R_4&-0.0030&0.0095&0.0090&0.0430&0.0541&0.0790&0.0757&0.0855\\ 
        R_5&0.0085&0.0442&0.0480&0.0541&0.0040&0.0090&0.0080&-0.0047\\ 
        R_6&0.0028&0.0085&0.0524&0.0790&0.0090&0.0440&0.0857&0.0845\\ 
        R_7&0.0025&0.0506&0.0075&0.0757&0.0080&0.0857&0.0410&0.0860\\ 
        R_8&-0.0536&0.0035&0.0088&0.0855&-0.0047&0.0845&0.0860&0.3200\\ 
        R_9&0.0785&0.0831&0.0853&-3.8168&0.0782&0.0422&0.0563&0.0001\\ 
        R_{10}&-0.0046&0.0065&3.9260&0.0452&0.0447&0.0467&0.2501&0.0548\\ 
        R_{11}&-0.0041&3.9240&0.0085&0.0431&0.0535&0.2500&0.0471&0.0431\\ 
        R_{12}&-3.9288&0.0066&0.0054&0&0.0085&0.0545&0.0436&0.0811\\ 
        R_{13}&-0.0017&0.0407&0.0550&0.2502&0.0080&0.0501&0.0459&0.0536\\ 
        R_{14}&-0.0678&-0.0017&0.0003&0.0545&-0.0026&0.0095&0.0529&0.0901\\ 
        R_{15}&-0.0543&0.0001&0.0014&0.0423&0.0063&0.0551&0.0105&0.0786\\ 
        R_{16}&-0.2499&-0.0538&-0.0687&-0.0011&-0.0581&-0.0048&0.0009&0.0860\\
        -&-&-&-&-&-&-&-&-\\
        &C_{9}&C_{10}&C_{11}&C_{12}&C_{13}&C_{14}&C_{15}&C_{16}\\
        \hline
        R_1&0.0785&-0.0046&-0.0041&-3.9288&-0.0017&-0.0678&-0.0543&-0.2499 \\
        R_2&0.0831&0.0065&3.9240&0.0066&0.0407&-0.0017&0.0001&-0.0538\\
        R_3&0.0853&3.9260&0.0085&0.0054& 0.0550&0.0003&0.0014&-0.0687\\
        R_4&-3.8168&0.0452&0.0431&0.0085&0.2502&0.0545&0.0423&-0.0011\\
        R_5&0.0782&0.0447&0.0535&0& 0.0080& -0.0026&0.0063&-0.0581\\
        R_6&0.0422&0.0467&0.2500&0.0545& 0.0501&0.0095&0.0551&-0.0048\\
        R_7&0.0563&0.2501&0.0471&0.0436&0.0459&0.0529&0.0105&0.0009\\
        R_8&0.0001&0.0548&0.0431&0.0811& 0.0536&0.0901&0.0786&0.0860\\
        R_9&0.3100&0.0845&0.0885&-0.0008&0.0830&0.0056&-0.0010&-0.0516\\
        R_{10}&0.0845&0.0390&0.0828&0.0080& 0.0759&0.0065&0.0526&-0.0010\\
        R_{11}&0.0885&0.0828&0.0410&0.0090& 0.0805&0.0541&0.0075&0.0028\\
        R_{12}&-0.0008&0.0080&0.0090&0.0050& 0.0529&0.0445&0.0437&0.0095\\
        R_{13}&0.0830&0.0759&0.0805&0.0529&0.0450&0.0070&0.0070&-0.0042\\
        R_{14}&0.0056&0.0065&0.0541&0.0445&0.0070&0.0040&0.0502&0.0075\\
        R_{15}&-0.0010&0.0526&0.0075&0.0437&0.0070&0.0502&0.0050&0.0090\\
        R_{16}&-0.0516&-0.0010&0.0028&0.0095&-0.0042&0.0075&0.0090&0.0440\\
        \end{array}} \right] + \nonumber
        \end{eqnarray}

\begin{eqnarray}
    i\left[{\begin{array}{c|cccccccc}
&C_{1}&C_{2}&C_{3}&C_{4}&C_{5}&C_{6}&C_{7}&C_{8}\\
    \hline
        R_{1}&0&-0.0248&-0.0130&-0.0411&0.0173&-0.0454&-0.0904&-3.3041\\
        R_{2}&0.0248&0&-0.0041&-0.0146&-0.0030&0.0158&-3.3054&0.0164\\
        R_{3}&0.0130&0.0041&0&0.0097&-0.0203&-3.3064&0.0193&-0.0806\\
        R_{4}&0.0411&0.0146&-0.0097&0&-3.1926&-0.0387&0.0020&-0.0562\\
        R_{5}&-0.0173&0.0030&0.0203&3.1926&0&-0.0283&-0.0140&-0.0839\\
        R_{6}&0.0454&-0.0158&3.3064&0.0387&0.0283&0&0.0011&-0.0941\\
        R_{7}&0.0904&3.3054&-0.0193&-0.0020&0.0140&-0.0011&0&-0.0668\\
        R_{8}&3.3041&-0.0164&0.0806&0.0562&0.0839&0.0941&0.0668&0\\
        R_{9}&0.0850&-0.0081&0.0030&-0.0294&0.0333&-0.0538&-0.0545&-0.2499\\
        R_{10}&0.0866&0.0095&0.0823&-0.0018&0.0556&0.0057&-0.0000&-0.0527\\
        R_{11}&0.0835&0.0265&0.0090&0.0021&0.0523&0.0000&0.0085&-0.0566\\
        R_{12}&0.0266&0.0448&0.0474&0.0090&0.2499&0.0538&0.0566&0.0333\\
        R_{13}&0.0527&0.0579&0.0533&-0.0004&0.0100&0.0074&0.0043&-0.0274\\
        R_{14}&0.0503&0.0257&0.2500&0.0540&0.0396&0.0055&0.0825&-0.0063\\
        R_{15}&0.0535&0.2500&0.0245&0.0542&0.0463&0.0267&0.0080&-0.0101\\
        R_{16}&-0.0001&0.0555&0.0553&0.0497&0.0286&0.0808&0.0901&0.0815\\
        -&-&-&-&-&-&-&-&-\\
        &C_{9}&C_{10}&C_{11}&C_{12}&C_{13}&C_{14}&C_{15}&C_{16}\\
        \hline
        R_1&-0.0850&-0.0866&-0.0835&-0.0266&-0.0527&-0.0503&-0.0535&0.0001\\
        R_2&0.0081&-0.0095&-0.0265&-0.0448&-0.0579&-0.0257&-0.2500&-0.0555\\
       R_3&-0.0030&-0.0823&-0.0090&-0.0474&-0.0533&-0.2500&-0.0245&-0.0553\\
       R_4&0.0294&0.0018&-0.0021&-0.0090&0.0004&-0.0540&-0.0542&-0.0497\\
       R_5&-0.0333&-0.0556&-0.0523&-0.2499&-0.0100&-0.0396&-0.0463&-0.0286\\
       R_6&0.0538&-0.0057&-0.0000&-0.0538&-0.0074&-0.0055&-0.0267&-0.0808\\
        R_7&0.0545&0.0000&-0.0085&-0.0566&-0.0043&-0.0825&-0.0080&-0.0901\\
        R_8&0.2499&0.0527&0.0566&-0.0333&0.0274&0.0063&0.0101&-0.0815\\
        R_9&0&-0.0623&-0.0993&-0.0913&-0.0612&-0.0793&-0.0496&-0.0573\\
        R_{10}&0.0623&0&0.0122&-0.0226&0.0016&0.0178&-0.0615&-0.0322\\
        R_{11}&0.0993&-0.0122&0&-0.0272&-0.0314&-0.0560&0.0193&-0.0497\\
        R_{12}&0.0913&0.0226&0.0272&0&0.0517&-0.0108&0.0019&0.0183\\
        R_{13}&0.0612&-0.0016&0.0314&-0.0517&0&0.0092&-0.0213&-0.0502\\
        R_{14}&0.0793&-0.0178&0.0560&0.0108&-0.0092&0&0.0053&-0.0241\\
        R_{15}&0.0496&0.0615&-0.0193&-0.0019&0.0213&-0.0053&0&-0.0258\\
        R_{16}&0.0573&0.0322&0.0497&-0.0183&0.0502&0.0241&0.0258&0\\
        \end{array}} \right ]
\end{eqnarray}
 
 where $R_i$ and $C_i$ denote the $i$th row and $i$th column of the matrix respectively.
\newpage 
\textbf{Circuit used for factorization of $N=175$}
\begin{figure}[H]
\includegraphics[width=\linewidth]{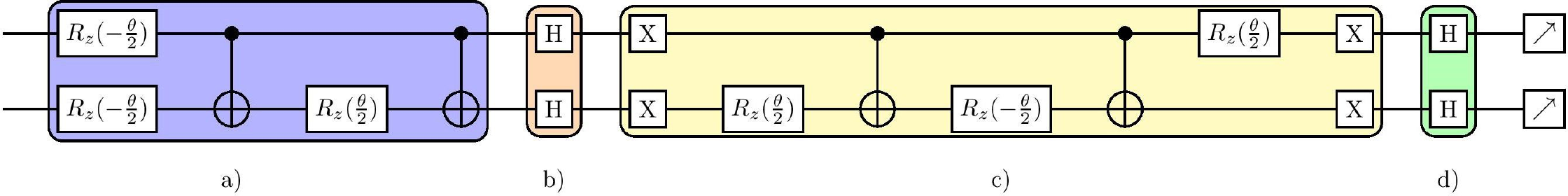}
\caption{\textbf{The circuit implementation of the protocol required for the factorization of $N=175$.} \textbf{(a)} $e^{i(\sigma_z\otimes I)\frac{\theta}{4}}$, $e^{i(I\otimes\sigma_z)\frac{\theta}{4}}$, $e^{-i(\sigma_z\otimes\sigma_z)\frac{\theta}{4}}$ are the components of $e^{-i\hat{H}\theta}$ operation. The $e^{-iI_{2}\frac{\theta}{4}}$ factor has been neglected. Overall, the action of this part of the circuit is that it conditionally induces a phase shift of $\theta$ angle only to the ground states of $\hat{H}$, which encode the solution to our problem, relative to the other basis states. Hence, our required ``solution" states have been marked as required for the quantum search algorithm. \textbf{(b)} Since our initial quantum database $\ket{\psi_0}$ from which the marked states are to be searched is taken to be the $2$-qubit equal superposition state, the operation $U$ such that $U\ket{00}=\ket{\psi_0}$ is given by $U=H^{\otimes2}$. In this step, we apply the $U^{\dag}$ operation, which is the corresponding inverse operation. In our case, $U^{\dag}$ is equal to $H^{\otimes2}$ as well. \textbf{(c)} As was the case in Fig.~\ref{qfa_sup_Fig1}, this part of the circuit is for the purpose of implementing the conditional phase shift $e^{i\theta}$ to $\ket{00}$ state only, while the other basis states are left unchanged. \textbf{(d)} Finally, we apply $U=H^{\otimes2}$ to obtain an equal superposition of the marked states only. It has been earlier discussed that the value of $\theta$ must be equal to $2sin^{-1}\left(\frac{sin\frac{\pi}{6}}{\frac{\sqrt{3}}{2}}\right)$ for the exact search algorithm to work. }
\label{qfa_sup_Fig3}
\end{figure}

%\subsection{Implementation in IBM Quantum Computer}

Fig.~\ref{qfa_sup_Fig3} presents the overall circuit for the given factorization problem. We performed the experiment $8192$ times in each requisite basis. The experimental density matrix was found to be (using Eq.~\eqref{qfa_eq_16})

\begin{equation}
   \rho^{E}= 
  \left[ {\begin{array}{cccc}
        0.3780 & 0.2430 & 0.2595 & 0.0010  \\
        0.2430 & 0.3520 & 0.2900 & -0.0375 \\
        0.2595 & 0.2900 & 0.2480 & -0.0945 \\
        0.0010 & -0.0375 & -0.0945 & 0.0220 \\
  \end{array} } \right]+
   i\left[ {\begin{array}{cccc}
   0.0000 & -0.1990 & -0.2625 & -0.2863\\
   0.1990 & 0.0000 & 0.0028 & 0.0390 \\
   0.2625 & -0.0028 & 0.0000 & 0.0530\\
   0.2863 & -0.0390 & -0.053 & 0.0000\\
  \end{array} } \right]
  \label{qfa_21}
  \end{equation}
  
The theoretically evaluated final state of the system is given by $\ket{\Psi}=\frac{1}{\sqrt{3}}\left[\ket{00}+\ket{01}+\ket{10}\right]$. Hence, the theoretical density matrix is given as
\begin{equation}
    \rho^{T}=\frac{1}{3} \left[ {\begin{array}{cccc}
        1 & 1 & 1 & 0  \\
        1 & 1 & 1 & 0 \\
        1 & 1 & 1 & 0 \\
        0 & 0 & 0 & 0 \\
    \end{array} }\right]
    \label{qfa_22}
\end{equation}

The state tomography has already been presented in the primary section of this paper.

\begin{figure}[H]
\centering
\includegraphics[scale=0.3]{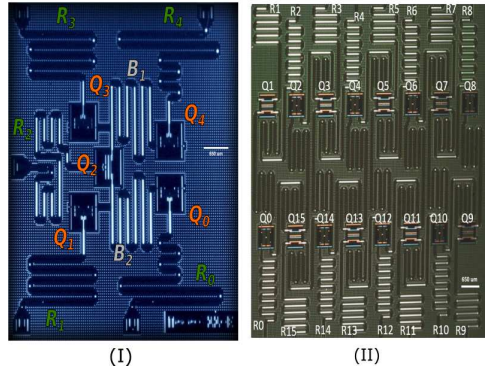}
\caption{\textbf{Layouts of the quantum processors used to carry out our experiments.} The figure illustrates the chip layout of 5-qubit quantum processor ibmqx4 (left) and 16-qubit quantum processor ibmqx5 (right). The chips are stored in a dilution refrigerator at temperature 0.021 K. \textbf{(I)} Here, all the 5 transmon qubits (charge qubits) are connected by two coplanar waveguide (CPW) resonators. The two CPWs couple Q2, Q3 and Q4 qubits with resonating frequency around 6.6 GHz and Q0, Q1 and Q2 qubits are coupled with 7.0 GHz frequency. Each qubit is controlled and readout by a particular CPW. The coupling map for the Control gates is represented as, $\{Q1 \rightarrow [Q0], Q2 \rightarrow [Q0, Q1, Q4], Q3 \rightarrow [Q2, Q4]\}$, where $a \rightarrow [b]$ means $a$ is the control qubit and $b$ is the target qubit for the implementation of Control gates. The gate and readout errors are of the order of $10^{-2}$ to $10^{-3}$. \textbf{(II)} Here the qubits are connected by total 22 coplanar waveguide (CPW) resonators, each of which connects two qubits. The coupling of Control gates is represented as ,$\{Q0 \rightarrow [Q1, Q15], Q2 \rightarrow [Q1,Q15, Q3], Q3 \rightarrow [Q2,Q14, Q4]\}$ and so on, where $a \rightarrow [b]$ means $a$ is the control qubit and $b$ is the target qubit.}
\label{qfa_sup_Fig4}
\end{figure}

\textbf{Experimental Setup \cite{IBM1,IBM2} :} The experimental parameters of `ibmqx4' chip are presented in Table \ref{tab1}, where $\omega^{R}_{i}$, $\omega_{i}$, $\delta_{i}$, $\chi$, $T_1$ and $T_2$ represent the resonance frequency, qubit frequency, anharmonicity, qubit-cavity coupling strength, relaxation time and coherence time respectively for the readout resonator. The connectivity and control of five superconducting qubits (Q0, Q1, Q2, Q3 and Q4) are depicted in Fig. \ref{qfa_sup_Fig4} (I). The single-qubit and two-qubit controls are provided by the coplanar wave guides (CPWs). The device is cooled in a dilution refrigerator at temperature $0.021$ K. The qubits are coupled via two superconducting CPWs, one coupling Q2, Q3 and Q4 and another one coupling Q0, Q1, Q2 with resonator frequencies $6.6$ GHz and $7.0$ GHz respectively. Individual qubits are used to to control and readout all the qubits. The connectivity on the $16$-qubit quantum processor `ibmqx5' is provided by total 22 coplanar waveguide (CPW) ``bus" resonators, each of which connects two qubits. Table~\ref{tab2} presents the experimental parameters of this processor, where $\omega^{R}_{i}$, $\omega_{i}$, $\delta_{i}$, $\chi$ and $\kappa$ represent the resonance frequency, qubit frequency, anharmonicity, qubit-cavity coupling strength and cavity coupling with the environment time respectively for the readout resonator. The fridge temperature is $0.0141929$ K. Three different resonant frequencies are used for the bus resonators, viz. $6.25$ GHz, $6.45$ GHz $6.65$ GHz. Each qubit has a dedicated CPW readout resonator attached (labelled as R) for control and readout. Fig.~\ref{qfa_sup_Fig4} (II) shows the chip layout.

\begin{table}[H]
\centering
\begin{tabular}{ c c c c c c c }
\hline
\hline
Qubits & $\omega^{R^{\bigstar}}_{i}/2\pi$ (GHz) & $\omega^{\dagger}_{i}/2\pi$ (GHz) & $\delta^{\ddagger}_{i}/2\pi$ (MHz) & $\chi^{\S}/2\pi$ (kHz) & $T^{||}_{1}$ ($\mu s$) & $T^{\perp}_{2}$ ($\mu s$)\\
\hline
\hline
Q0 & 6.52396 & 5.2461 & -330.1 & 410 & 35.2 & 38.1 \\
Q1 & 6.48078 & 5.3025 & -329.7 & 512 & 57.5 & 40.5 \\
Q2 & 6.43875 & 5.3025 & -329.7 & 408 & 36.6 &54.8 \\ 
Q3 & 6.58036 & 5.4317 & -327.9 & 434 & 43.0 & 42.1 \\
Q4 & 6.52698 & 5.1824 & -332.5 & 458 & 49.5 & 19.2\\
\hline
\hline
\end{tabular}\\
$\bigstar$ Resonance frequency, $\dagger$ Qubit frequency, $\ddagger$ Anharmonicity, $\S$ Qubit-cavity coupling strength, $||$ Relaxation time, $\perp$ Coherence time.
\caption{\textbf{The table shows the parameters of the device ibmqx4.}}
\label{tab1}
\end{table} 

\begin{table}[H]
\centering
\begin{tabular}{ c c c c c c }
\hline
\hline
Qubits & $\omega^{R^{\bigstar}}_{i}/2\pi$ (GHz) & $\omega^{\dagger}_{i}/2\pi$ (GHz) & $\delta^{\ddagger}_{i}/2\pi$ (MHz) & $\chi^{\S}/2\pi$ (kHz) & $\kappa^{||}/2\pi$\\
\hline
\hline
Q0&	6.9745&	5.2561&	-292.8&	125&    345 \\
Q1&	6.8667&	5.3961&	-291.0&	153&	373 \\
Q2&	6.96087&	5.2756&	-289.2&	118&	440 \\
Q3&	6.87813&	5.0831&	-294.0&	100&	345 \\
Q4&	6.95278&	4.9791&	-290.7&	81&	545 \\
Q5&	6.85144&	5.1513&	-286.0&	125&	372 \\
Q6&	6.98235&	5.3058&	-289.2&	127&	413 \\
Q7&	6.85953&	5.2528&	-298.4&	128&	349 \\
Q8&	6.97061&	5.1153&	-287.6&	120&	321 \\
Q9&	6.87244&	5.1555&	-297.6&	72&	299 \\
Q10&	6.95882&	5.0426&	-291.5&	83&	428 \\
Q11&	6.87644&	5.1107&	-290.7&	100&	561 \\
Q12&	6.96631&	4.9466&	-299.0&	81&	550 \\
Q13&	6.86321&	5.0881&	-289.8&	108&	398 \\
Q14&	6.97820&	4.8701&	-296.0&	78&	545 \\
Q15&	6.85637&	5.1095&	-289.8&	109&	471 \\
\hline
\hline
\end{tabular}\\
$\bigstar$ Resonance frequency, $\dagger$ Qubit frequency, $\ddagger$ Anharmonicity, $\S$ Qubit-cavity coupling strength, $||$ Cavity coupling to the environment.
\caption{\textbf{The table shows the parameters of the device ibmqx5.}}
\label{tab2}
\end{table}

%\section{Discussion and Conclusion}

%\pagebreak

%\onecolumngrid
%\section*{ Supplementary Information: Hardware-efficient Quantum Optimizer for Small Molecules and Quantum Magnets}
%\beginsupplement
%\section{Device and characterization}

\end{document}